\documentclass[Journal]{IEEEtran}
\usepackage{cite}
\usepackage{epsfig,graphics,mathrsfs,amssymb,amsmath,bm,color,yfonts,txfonts,multirow,multicol,slashbox,dblfloatfix}
\usepackage{subfigure}
\usepackage{algorithm2e}
\usepackage[noend]{algpseudocode}

\begin{document}

\title{Next Generation New Radio Small Cell Enhancement: Architectural Options, Functionality and Performance Aspects}
\author{Mirza Golam Kibria, Kien Nguyen, Gabriel Porto Villardi, Kentaro Ishizu and Fumihide Kojima
}
\maketitle

\begin{abstract}

The 3rd generation partnership project (3GPP) has been engaged in further advancing the evolved universal mobile telecommunications system (UMTS) terrestrial radio access network (E-UTRAN) and UTRAN based radio access network technologies. New radio (NR) is the 3rd generation partnership project (3GPP) endeavor for outlining and standardization of the 5th generation (5G) advanced radio access technology. 3GPP has released the first set of 5G NR standards, i.e., the non-standalone 5G radio specifications. As long-term evolution (LTE) technology is massively deployed and broadly accepted, the transition from LTE to 5G is very critical, and it is of maximal importance that the backward compatibility of 5G with LTE is considered. 3GPP has identified several architecture options for 5G. This article gives an overview of the NR architecture options, their deployment scenarios, and the key migration paths. The LTE-NR dual connectivity (DC) is presented. This DC scenario is unique in the sense that DC is being endowed for two different generations of 3GPP radio access technologies. We, further, present the integration of multipath transmission control protocol (MPTCP) with LTE-NR DC and DC-like aggregation, i.e., 3GPP-non-3GPP interworking to bring in the advantages of MPTCP in terms of link robustness, reliability and dynamic mapping between the traffic flows and the available paths. Finally, we discuss the future research and standardization directions of the next-generation networks.

\end{abstract}

\IEEEpeerreviewmaketitle
\section{Introduction}

New radio (NR) is the 3rd generation partnership project (3GPP) endeavor for outlining and standardizing the advanced radio access technology for 5th generation (5G)\cite{Ref1}. NR is a longstanding effort embracing a large set of use cases such as enhanced mobile broadband (eMBB), ultra-reliable and low-latency communications, massive machine type communications, and opening up several new leading-edge technologies like mmWave and three-dimensional beamforming. Wireless and mobile data appetency continues to grow rapidly. New device classes, for instance, virtual and augmented reality headsets, cloud-based artificial intelligence (AI)-enabled devices, connected and autonomous cars, are gaining traction and poised to make use of the new 5G infrastructure and capabilities. 

The 3GPP has identified several architecture options for 5G \cite{3GPP5}. Mobile network operators (MNOs) are accelerating their network deployment and/or upgrade plans. The full-scale roll out of 5G NR standalone system is the long-time final goal. It is expected that majority of the MNOs will have one or more intermediate states for quite a long time. Over the migration period, the current long-term evolution (LTE)/LTE-advanced (LTE-A) is expected to deliver general extended coverage and mobility, and the NR is anticipated to facilitate boosting of user data capacity when and where the traffic load is high. NR at both sub-6 GHz and mmWave as well as legacy cellular formats will be supported by the 5G modem solutions.

Deployment of small cells delivers expansive coverage, enhanced throughput, and mitigates the massive traffic burden in macro-cells.
A macro base station (BS) provides services over a large coverage area while a micro BS serves relatively a small area within the macro coverage. The 3GPP Rel-12 has specified the dual connectivity (DC)\cite{Ref3} feature that allows a UE to communicate simultaneously through both macro BS and small cell BS. The DC can considerably enhance the data throughput especially for cell-edge UEs, mobility robustness and reduce the signaling overhead towards the core network (CN).
\begin{figure*}
  \centering
   \includegraphics[scale=.5]{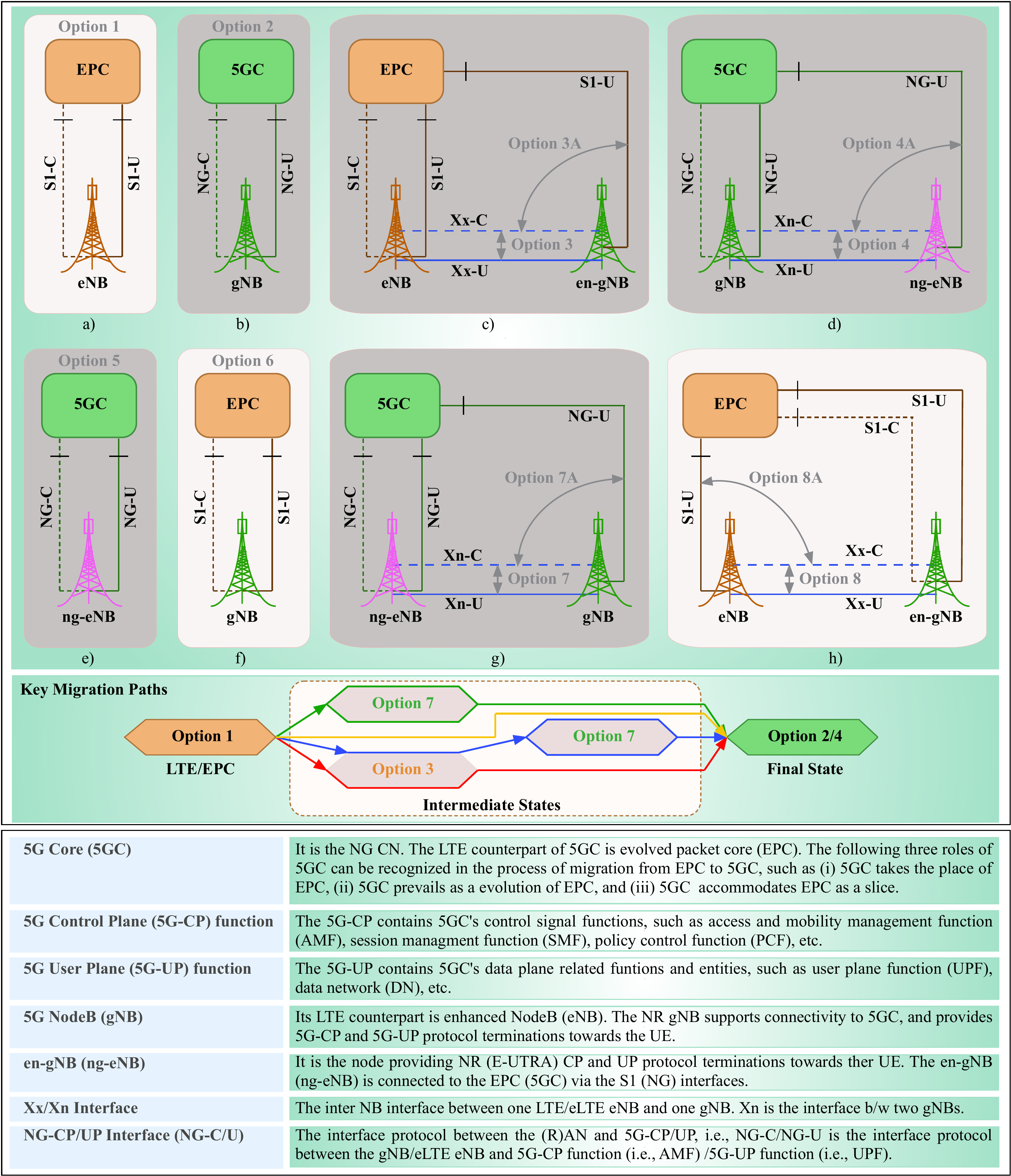}
   \caption{5G architectural options recognized by 3GPP, and short introduction to introduced network entities and/or functions and interfaces for NR. The key migration paths are also shown. Lines with different colors represent different migration paths. Note that gNB logical architecture consists of a central unit (CU) and distributed unit (DU), i.e., gNB=CU+DU. The CU controls the functioning of the DUs over the front-haul interface.}
   \label{fig_NR_1}
\end{figure*}
Small cell enhancement through DC under different 5G architecture options, especially under the interworking between NR and (evolved) E-UTRA needs to be explored.

Networks are multi-path: user equipments (UEs), in general, have multiple network interfaces, for example, the cellular interface, the Wi-Fi interface, etc. Since an application's data can be delivered to the destination through multiple connections, out of order delivery and multiple latencies need to be supported in DC. Nonetheless, reliable communication scenarios have not been addressed in current 3GPP legacy DC architecture. Robust and/or reliable communications/applications, in general, involve bearer duplication and/or backup across multiple links. Based on legacy network architecture, the system needs to incorporate features related to the duplication process that would certainly result in added complexity. Furthermore, in order to benefit the most from the DC feature, dynamic selection of the most suitable path for a given bearer, and at the same time, loading/unloading of less/more congested paths are very important. One of the best solutions that are able to deal with such cases is multi-path transmission control protocol (MPTCP)\cite{Ref7}. Although the current protocol stack in 3GPP offers the structure and support for MPTCP, associating such multi-path flows to DC is quite missing.

This article discusses the 5G NR architecture options, their deployment scenarios, and the key migration paths. We present the MPTCP over DC schemes that bring in the advantages of MPTCP in 3GPP cellular in terms of link robustness, reliability and dynamic mapping between the traffic flows and the available paths. It also gives an overview of variants of DC, i.e., DC-like aggregation, e.g., multi-access connectivity, interworking of 3GPP and non-3GPP access along with MPTCP under NR architecture options.

\section{The 5G Architecture Options}
A total of twelve architectural options have been recognized, which are shown in Fig.~\ref{fig_NR_1}. Note that the architectural options identified encompass all potential deployment scenarios starting from the legacy LTE to full-fledged 5G. However, not all the architectural options will be practically implemented. The options with the darker background, i.e., Options 2, Option 3/3A, Option 4/4A, Option 5 and Option 7/7A are the most fitting solutions for delivering NR access to capable UEs. Before we discuss the options in detail, from Fig.~\ref{fig_NR_1}: up note that the new (radio) access network ((R)AN) and CN consist of several new logical entities and interfaces. We give a short introduction to the newly introduced network functions and interfaces and some of the network functions in Fig.~\ref{fig_NR_1}: down. The 5G system (5GS) architecture consists of a large number of network functions. The functional descriptions of the network functions, the interworking between network functions, the point-to-point (i.e., pair-wise) reference points, i.e., the interfaces connecting the network functions (e.g., (R)AN internal interfaces and CN internal interfaces, (R)AN-CN interfaces) and the service-based interfaces are provided in 3GPP TS 23.501 V15.0.0.

In the following, we briefly discuss the most likely solutions to be practically implemented to deliver NR access. Note that Option 1 is legacy LTE system. Note that the 5G architecture options are identified according to different 5G deployment scenarios and probable migrations paths. Most of the operators may not migrate all of the LTE deployments to 5G overnight because of the huge cost, inter-system backward/forward compatibility issues involved in it. Multiple architecture options are available depending on (i) different combinations of CN (EPC and 5GC), (ii) whether the system is SA (LTE, 5G) or NSA (LTE eNB anchor, NR gNB anchor).
\begin{itemize}
\renewcommand{\labelitemi}{\scriptsize$\blacksquare$}
\item {\bf Option 2: Standalone (SA) NR in 5GS.} This deployment scenario is specifically attractive in areas where there is no legacy LTE system and full-fledged 5G NR access system is required to be deployed. For early deployment of full-fledged 5G networks, this is the most attractive option where the operators can introduce 5G-only service without 4G interworking. In SA NR, the gNB connects to the 5GC. The full-suite of 5G specifications, for 3GPP Rel-15 (i.e., 5G Phase-1), will define the SA NR system.
\item {\bf Option 3/3A: Non-Standalone (NSA) NR in Evolved Packet System.} Under this architectural option, 5G NR will utilize the existing LTE radio and CN as an anchor for mobility management and coverage while adding a new 5G carrier. This option is, in particular, attractive for early deployments of 5G NR access systems  in areas where legacy eNB and EPC are operational. It is attractive to many MNOs as it does not need a 5GC. The anchor LTE eNB is connected to the EPC, and the NR UP (part of the network carrying user data) connection to the 5GC goes through the LTE eNB (Option 3) or directly (Option 3A). 3GPP has released the NSA 5G radio specifications.
\item {\bf Option 4/4A: NSA Evolved E-UTRA in 5GS.} This deployment scenario is especially attractive for deployments of NR access systems in areas where legacy LTE eNB and the EPC are prepared/qualified to be upgraded to ng-eNB and the 5GC, respectively, in order to inherit the benefits of these enhanced network elements. Under Option 4/4A, the gNB is connected to the 5GC with NSA evolved E-UTRA, and the ng-eNB needs the NR gNB as an anchor for CP (part of the network carrying singling traffic) connectivity to 5GC. The evolved E-UTRA UP connection to the 5GC goes through the gNB (Option 4) or directly (Option 4A). 
\item {\bf Option 5: SA Evolved E-UTRA in 5GS.} This deployment scenario is especially fitting in areas where there is no legacy LTE system and evolved E-UTRA access systems are deployed. Under this deployment scenario, the ng-eNB is connected to the 5GC. 
\item {\bf Option 7/7A: NSA NR in 5GS.} This option is, in particular, attractive for areas where legacy LTE eNBs and the EPCs are prepared/qualified to be upgraded to ng-eNBs and the 5GCs. This is NSA from the point of view of the gNB, which requires an ng-eNB as an anchor for CP connectivity to 5GC. Here, the ng-eNB is connected to the 5GC, and the NR UP connection to the 5GC goes through the ng-eNB (Option 7) or directly (Option 7A).
\end{itemize}

\subsection{Migration: 4G to 5G}
The deployment of 5GS without interworking between EPC and 5GC is the long-time final goal after migration under the assumption that EPC may still last for a long time to provide legacy or roaming UE support. Therefore, Option 2/Option 4 can be considered as the target architecture but for the majority of the MNOs may involve one or more intermediate states (e.g., Option 3 or Option 7) in some cases over a longer period of time. The requirements of the 5GS should not be compromised or sacrificed in order to fit a specific intermediate state or migration paths. Four key migration paths have been identified in \cite{RefX} which are graphically shown in Fig.~\ref{fig_NR_1}: up. For more details on the migration paths, migration time and the associated cost, please refer to \cite{RefX,RefY}.

A critical feature of the 5GC strategy for most of the MNOs is to efficaciously handle the migration from EPC to 5GC.  Separation of CP and UP processing is an integral part of the 5GS. Introducing control- and user-plane separation (CUPS) to EPC yields a worthwhile migration passage from 4G to 5G. A logical mapping amidst the LTE/LTE-A and 5G architectures that yields a hybrid EPC/5GC (i.e., serving both network types) can also be exploited in CN migration\cite{Lightreading}. For example, in UP, a converged gateway holding up both UPF and S/PGW-U can be used. Likewise, in CP, PGW-C, PCRF, and HSS of LTE can be combined with SMF, PCF, and UDM, respectively.

\subsection{Challenges}
The NR Opportunities come with various challenges. The MNOs must outplay various crucial challenges in technology advancement and revolution to unchain the 5G potential. For example, mmWave propagation and channel modeling, protocol optimization, antenna complexity, digital interface capacity are some of the challenges. 
Development of interworking functionality (between EPC and 5GC) is very critical. One of the main challenges is to develop interworking functionality with minimal interfaces between the EPC and the 5GC so that 5GC has no longer dependency on the interworking. As such the (i) EPC enhancement in terms of extending the range of quality of service (QoS), upgrading the UE capabilities, resolving UE compatibility issues and handling the UE mobility issues to support 5G NR via DC (e.g., Option 3), and (ii) legacy LTE system enhancement for EPC connectivity to 5GC in terms of supporting slicing, mobility between LTE and NR (both connected to 5GC, e.g., Option 7), yield immense challenges. Executing new potentialities of NR also imposes some challenges given new functions like flexible air-interface, channels codes, active antenna systems, etc. More key issues are discussed in \cite{RefY}.

\section{Dual Connectivity}

The DC feature allows UE to have two independent connections to master node (MN)\footnote{In legacy LTE, it is denoted as master eNB (MeNB).} (the BS that terminates at least the CP and serves as mobility anchor towards the CN), and a secondary node (SN) (a node different from the MN that delivers added/supplementary radio resources to the UE), simultaneously. A macro/small cell BS would typically be the MN/SN. Note that the BSs/NBs are more likely to take different roles for different UEs, for example, an MN to one particular UE can act as the SN or the only BS to another UE. A master cell group (MCG)/secondary cell group (SCG) is defined as a group of serving cells associated with the MN/SN.

\begin{figure*}
  \centering
   \includegraphics[scale=.5]{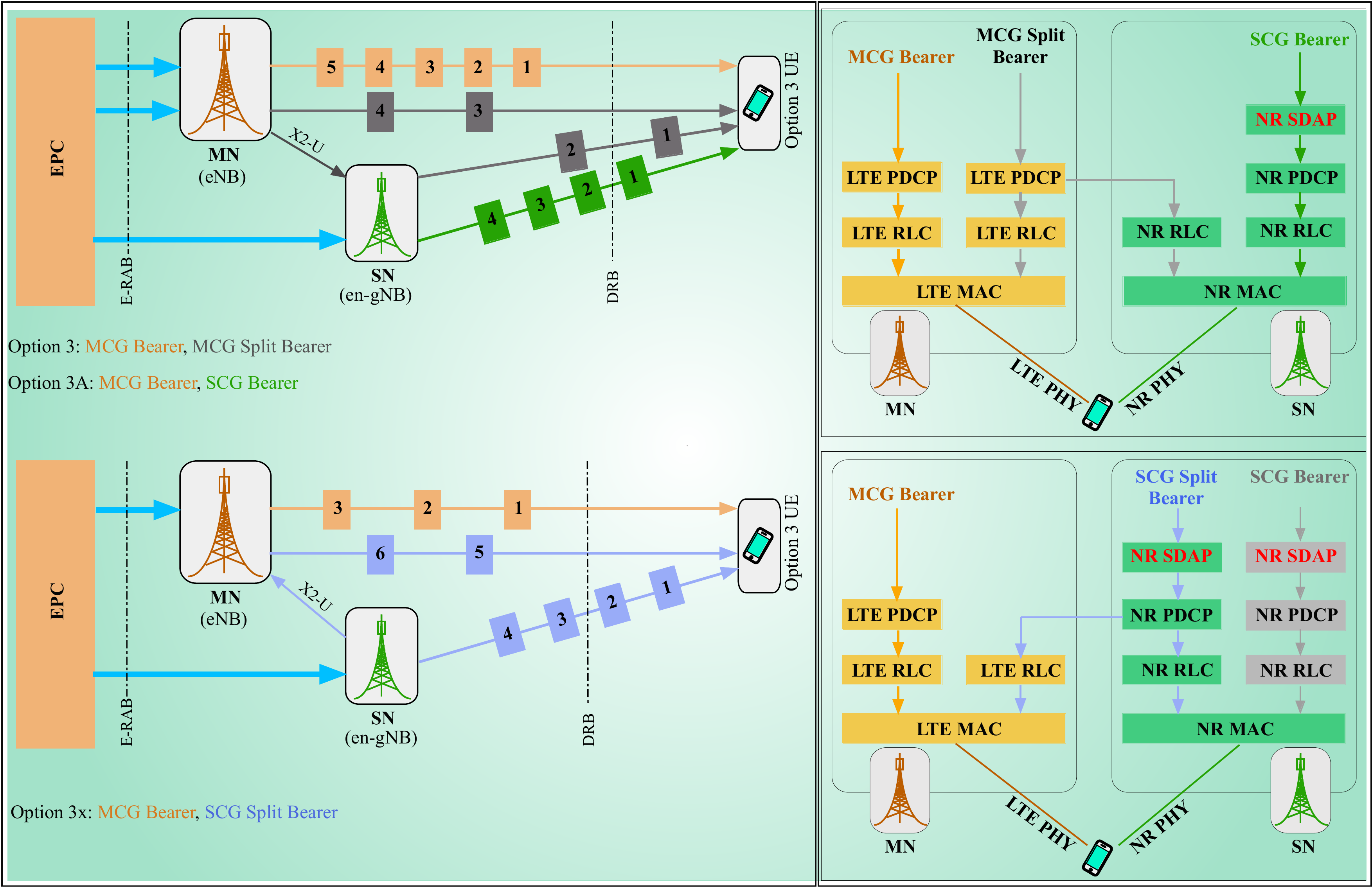}
   \caption{Left: Bearers in EN-DC. Option 3: Packet level split at SGW, Option 3A: bearer level split at MN, Option 3x: bearer level split at SN (Split bearer enhanced introduced in NR). Note that MCG Split Bearer and SCG Split Bearer as well as MCG Split Bearer and SCG Bearer cannot be configured simultaneously. MCG/SCG (split) bearers exist between MN/SN and UE only. The DRB terminate in the MN/SN and UE. E-UTRAN radio access bearer (E-RAB) is established between EPC and gNB. Right: The radio protocol architecture of the MN (left) and the SN (right). The MN and the SN use their own PHY techniques, i.e., SN can employ the NR PHY techniques such as mmWave, massive MIMO, dynamic beamforming. Note that in EN-DC, the UE has a second RRC  termination at the SN, unlike the LTE DC, to trigger intra-NR mobility.}
   \label{fig_NR_2}
\end{figure*}

\subsection{DC under interworking between (evolved) E-UTRAN and NR}

To facilitate effective interworking between the NR and the (evolved) E-UTRA, a data-flows aggregation technology based on LTE DC is investigated in \cite{Ref5} while a technology of aggregating NR carriers is studied \cite{Ref1}. In general, both NR gNB and (e)LTE eNB can serve as the MN. However, DC solution with (e)LTE eNB as the MN will initially be prioritized, and later on, NR gNB can be the MN or works as standalone BS. Note that Option 3/3A, 4/4A, and 7/7A can be considered a tight interworking between NR and (evolved) E- UTRA defined under multi-radio access technology DC (MR-DC) architecture.

\begin{itemize}
\renewcommand{\labelitemi}{\scriptsize$\blacksquare$}
\item {\bf E-UTRA-NR DC (EN-DC) Under Option 3/3A}: It is the first phase of 5G (3GPP Rel-15) DC scheme and can only be operated with an LTE-Advanced Pro (3GPP Rel-13/Rel-14) eNB. 
Since EPC is employed as the CN, DC procedures specified in \cite{Ref5} and conforming stage 3 specifications in 3GPP TS 36.423 V15.0.0 can be utilized. Moreover, the protocols and procedures of the interface Xx between MN and SN will most likely be similar to \cite{Ref5} while there can be some insignificant enhancements.

\item {\bf NG-(R)AN Supported NR-E-UTRA DC (NE-DC) Under Option 4/4A}: The gNB takes the role of the MN and ng-eNB serves as SN \cite{3GPP10}. The role of the MN/SN is similar to the role of MeNB/SeNB in 3GPP TS 36.300. The protocols and procedures of the interface between MN and SN need to be newly defined.

\item {\bf NG-(R)AN Supported NG-(R)AN E-UTRA-NR DC (NGEN-DC) Under Option 7/7A}: In this DC scheme, the ng-eNB takes the role of an MN and connects to the 5GC with NSA NR, where the gNB acts as SN. The role of the MN/SN is similar to the role of MeNB/SeNB in legacy DC\cite{Ref5}. The NR gNB serves as a supplementary carrier. The protocols and procedures of the interface between MN and SN need to be outlined.
\end{itemize}

\subsection{Bearer for DC Between 5G NR and E-UTRA}
 
 In legacy LTE DC, a bearer is a virtual connection between the UE and public data network (PDN) gateway (PGW) that transports data with specific QoS attributes. However, the bearer concept is not considered in the 5GC network while NR is expected to maintain the radio bearer concept. A radio bearer can be a signaling radio bearer (SRB) (bearer of CP data) and/or data radio bearer (DRB) (bearer of UP internet protocol (IP) packets). Note that there are three different bearers defined for legacy E-UTRA DC, such as MCG bearer, SCG bearer and MCG Split bearer, which are still maintained in the 5GS. The MCG/SCG bearer uses the resources in MN/SN and follows the radio protocols only located in MN/SN. MCG Split bearer uses the radio protocols located in both MN and SN.
 When MCG Split bearer is employed in downlink, the packet routing decision is taken by the MN, and the decision relies on several parameters such as channel conditions, traffic load, buffer status, non-ideal backhaul capacity.

{\it {NR Bearer Enhancement: SCG Bearer Split:}} 
The 5G NR embraces SCG Split bearer in DC\cite{3GPP5}, and to make it work, deployment options different from the ones already discussed need to be supported. For example, under the NSA NR deployment scenario, when the MN (LTE eNB) is connected to EPC, the bearer of the UP data from EPC is split at the NR gNB. This new deployment option is referred to as 3x. Under this deployment option, the CP is still terminated at the legacy MN. Similarly, when the ng-eNB is connected to the 5GC with NSA NR gNB, the UP data over the NG-U interface is split at the NR gNB. This newly defined deployment option is referred to as 7x, where the NG-C interface carrying CP signal is terminated at the MN.
Splitting at the MN/SN supports both bearer level and packet level splitting. The left part of Fig.~\ref{fig_NR_2} shows all bearer types and the places where splitting may take place.

The radio protocol architecture of the MN, SN and the UE depends on how the radio bearer is set up and the deployment scenario. The MN/SN radio protocol architecture for deployment options 3, 3A and 3x are shown at the right end of Fig.~\ref{fig_NR_2}. The radio protocol architecture for the MCG, SCG and the Split Bearers from a UE perspective in MR-DC with EPC (EN-DC) and MR-DC with 5GC (NGEN-DC, NE-DC) are provided in \cite{3GPP10}. It is worth mentioning that in NR, a new access stratum sublayer, service data adaptation protocol (SDAP) sublayer, is introduced above the PDCP. The fundamental functions of the SDAP sublayer include: (i) mapping between a DRB and a QoS and (ii) marking QoS identification in data packets.

\begin{figure*}
  \centering
   \includegraphics[scale=.5]{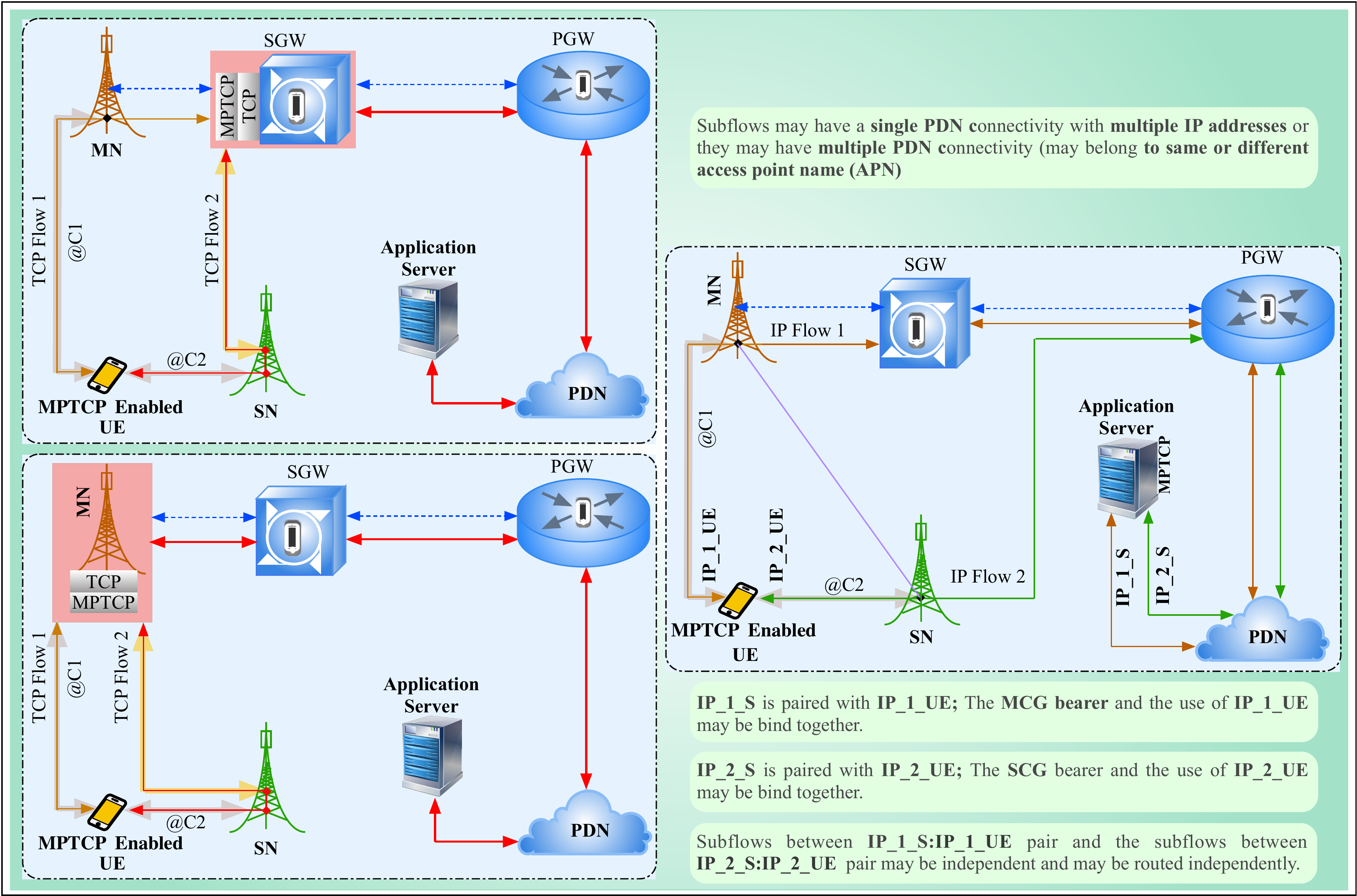}
   \caption{Left: MPTCP over DC under NR deployment scenario 3/3A. The serving gateway (SGW) includes an MPTCP function (up); Anchor eNodeB may terminate downlink TCP flows and initiate MPTCP sub-flows (down). Right: MPTCP over NR DC with two independent network interfaces at the UE. The IP flows from a single PDN connection use different IPv6 prefixes, which are independently routable. It is an end-to end MPTCP deployment scenario, and may considered to be inter NR-evolved E-UTRA DC option 3A as integrating the SGW with the PGW is a recognized implementation. }
   \label{fig_NR_3}
\end{figure*}

\section{MPTCP over EN-DC}

The MPTCP architecture is generally defined for two or more independent network interfaces, i.e., IP connections. However, in DC, all the bearers associated with a PDN share the same IP address of the UE. In DC, the presence of multiple data-link and physical carriers is covert from the IP layer, and as a result, both the IP and transport layer are aware of only a single network interface. However, in order to employ MPTCP over DC, the upper layer such as network layer and the transport layer need to be aware of the existence of multiple data-link layers and multiple carriers. Consequently, to incorporate performance advantages of MPTCP in DC, the system should have some mechanisms to make the upper layer aware of the presence of multiple physical carriers. A mechanism to use MPTCP over DC when only a single IP interface is devised in \cite{Patent1}, i.e., only one IP address is instantiated at the UE as shown in Fig.~\ref{fig_NR_3}. The mechanism is based on a new or extended application programming interface (API) signaling between different layers in order to make the upper layers aware of the existence of multiple data-link and physical carriers, and mapping between the data-paths and the data-link layers. 

When the SGW includes the MPTCP function, it terminates the downlink TCP flows and initiates MPTCP sub-flows towards the MN and the SN. Alternatively, the anchor MN can also terminate the downlink TCP flows and initiate MPTCP sub-flows. In DC with MPTCP feature, the MNO/SGW decides which sub-flows are transmitted over which component carrier based on radio conditions, delay characteristics of the carriers, packet loss characteristics of the available links, cost of transmitting a packet over one carrier versus another, packet QoS. These MPTCP over DC configurations offer different types of operations and have their own advantages and disadvantages according to characteristics of the SCG bearer and MCG Split bearer. When MPTCP is terminated at the SGW, there is one S1 (S1-U+S1-C) interface (between EPC and MN) and one 5G (NG-U +NG-C) interface (between EPC and SN). As a result, the load is distributed. But when MPTCP is terminated at the MN, there is only one S1 interface and as a result, the S1 gets overburdened, especially when the MN acts as the anchor for many UEs. By terminating MPTCP at the SGW, the loading balancing in DC can be done at EPC rather than at MN.

The network elements/functions will be affected by these different MPTCP over DC deployment scenarios. For example, when MPTCP is terminated at the SGW, the MN does not need to buffer or process packets of the SN bearer, but the SN mobility becomes apparent to CN. As a result, when the UE moves to a different SN, there yields the handover-like interruption period. On the other hand, when the MPTCP is terminated at the MN, unlike the previous scenario, there is no interruption period as the PDUs can be steered via MN at SN change, and dynamic reconfiguration is performed at RAN level. However, the MN needs to buffer and process the packets of the MCG and need to route traffic via MN. Therefore, there have been some trade-offs between these MPTCP over DC deployments.

The benefits of MPTCP over DC can be further enhanced if the bearers can have different routing paths between the UE and application server as MPTCP is mainly designed to work with multiple IP interfaces. In order to extract the full performance gain from the MPTCP feature, the UE may be allowed to use different IPv6 addresses for the available data paths, but within a single PDN connection, as in legacy DC. The Internet Protocol version 6 (IPv6) multi-homing approach\cite{Patent2, Patent3} allows the UE and the PGW to use multiple IPv6 prefixes over a single PDN. As such, the MPTCP enabled UE can request from a single PGW two differently routable IP addresses. Unlike the conventional DC where both the bearers have the same IP address, in this configuration, the bearers have different IPv6 prefixes, which are independently routable as shown in Fig.~\ref{fig_NR_3}: right. The MPTCP feature at the UE/Application Server may perceive the data path properties from the fixed network transport paths characteristics as well as from the link behavior of the MN and the SN. Since the MPTCP resides in the end-nodes, MPTCP/DC mapping is not required to be aware of DC/MPTCP mapping.

\begin{figure*}
  \centering
   \includegraphics[scale=.5]{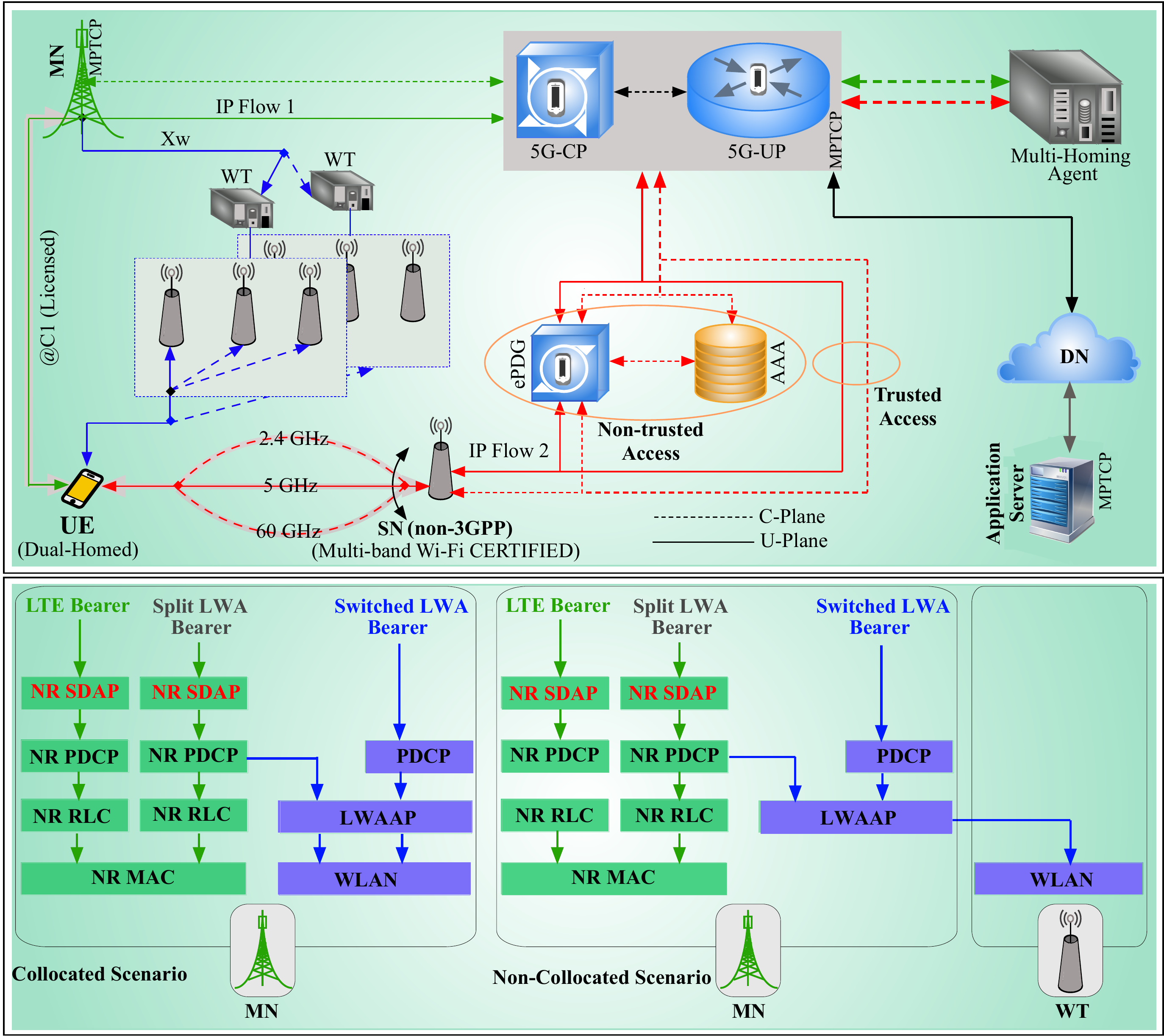}
   \caption{Up: Schematic for multi-access connectivity with NR. The entities/functions connected by the green and red lines belong to the legacy 3GPP-non-3GPP interworking feature, and the entities/functions connected by the green and blues lines correspond to eLWA feature. Down: Protocol stacks for both collocated and non-collocated eLWA. The interworking between ePDG/EPC, 3GPP AAA server and 5GS is studied in 3GPP TS 23.501 V15.0.0.}
   \label{fig_NR_4}
\end{figure*}

\subsection{Multi-Access Connectivity and MPTCP}
\label{MADC}

The ever-increasing mobile data traffic creates difficult challenges for the operators, especially when their licensed spectrum is limited.
The non-3GPP access networks, such as wireless fidelity (Wi-Fi), WiMAX are widely deployed at home, office and through various hot-spots. The Wi-Fi interface is also collocated with 3GPP based interfaces, e.g., LTE/LTE-A, in most devices. As a result, it facilitates the operators to benefit from seamless traffic offloading to Wi-Fi, i.e., wireless local area network (WLAN) as well as carrier aggregation (both licensed and unlicensed) or opportunistic use of unlicensed spectrum. Traffic offloading to unlicensed and/or shared spectrum helps the operators to better manage the available spectrum in the presence of greedy users, thus maximize the utilization of all available resources.

{\it \bf 3GPP 5G NR-non-3GPP interworking:} 3GPP TR 23.861 V13.0.0 allows the UEs to have multi-access connectivity, where the UE is capable of using multiple radio interfaces at a time. The UE can be connected to 3GPP and WLAN, MulteFire or WiMAX access network simultaneously in the CN gateway. However, it is common that the 3GPP network and the WLAN access network use different IP addresses as they, in general, belong to different APN. Fortunately, there are several existing techniques that preserve the same IP for delivering IP flows between 3GPP and WLAN access network.

The IP from mobility (IFOM)\cite{Oliva} feature facilitates seamless traffic or IP flows offloading from 3GPP network to WLAN access network. By virtue inter-system routing policy of IFOM, IP flows from a single PDN can be transferred via different and selected access networks. Therefore, IFOM and MPTCP together facilitate the UE to have two different active access networks under a single PDN connection, and the IP flows can be dynamically moved between the available different access networks. Using IFOM, the QoS demanding applications can be served via the 3GPP access while the best-effort traffic can be routed via the WLAN access network. A multi-homing agent placed in the CN can also play the very important supporting role in multi-access DC over MPTCP.

The non-3GPP access network connection with 3GPP CN \cite{3GPP9} is classified as either trusted or non-trusted as shown in Fig.~\ref{fig_NR_4}. The trusted non-3GPP access network is directly connected to the PGW, while the non-trusted non-3GPP access network is connected to the PGW through evolved packet data gateway (ePDG). The 3GPP authentication, authorization and accounting (AAA) server provides additional authentication and security checks in order to allow non-3GPP access to 3GPP CN. A multi-band Wi-Fi CERTIFIED access network supporting 2.4 GHz, 5 GHz and 60 GHz (WiGig) bands can support hand-off between the frequency bands, thus allows the selection of the most suitable band and data-rate for the application and the channel conditions as shown in Fig.~\ref{fig_NR_4}.

{\it \bf NR-Wi-Fi interworking:} Note that under the legacy 3GPP-non-3GPP interworking, the carrier needs to deploy and maintain two separate networks. Moreover, the interworking does not offer the bearer (a single bearer) splitting function. Enhanced aggregation LTE-WLAN (eLWA) is a 3GPP Rel-14 feature, which allows a UE to utilize network's LTE and Wi-Fi links simultaneously. Unlike legacy LTE-WLAN interworking, the data in the Wi-Fi link under eLWA originates in the (e)LTE eNB/NR gNB (if SA NR is deployed). Again, unlike the legacy 3GPP-non-3GPP access, WLAN does not interact with the CN under eLWA. In eLWA, the WLAN is fully controlled by the anchor node (i.e., the cellular node acts as CP and UP anchor), thus simplifies the WLAN integration with the cellular network. The eLWA feature is based on 3GPP DC like framework, and as a result, eLWA can utilize parts of DC functionality. The non-collocated eLWA architecture ((e)LTE eNB/NR gNB and the WLAN termination (WT) are not integrated) is depicted in Fig.~\ref{fig_NR_4}: up. 

The eLWA architecture eliminates costly WLAN-specific dedicated core network ePDG by integrating the WLAN at the RAN level. As already mentioned, the NR gNB can schedule the PDCP PDUs belonging to the same bearer to be delivered to the UE either through the WLAN or NR, and aggregation of NR and WLAN occurs at the PDCP level. The protocol stacks for both collocated and non-collocated eLWA are shown in  Fig.~\ref{fig_NR_4} (bottom). Similar to DC, there are three different bearers in eLWA, which are NR bearer, Split LWA bearer (split between NR and WLAN) and switched LWA bearer. The PDCP PDUs transported through WLAN are encapsulated in LWA adaptation protocol (LWAAP), which carries bearer identity. Note that end-to-end (UE-to-Application Sever) or end-to-middlebox (UE-to-5G-UP/NR gNB) MPTCP deployment can also be performed in multi-access DC scenarios. End-to-end MPTCP facilitates faster handover between the non-3GPP and 3GPP networks.

\subsection{Cost and Feasibility of MPTCP Deployment}
The feasibility of MPTCP deployment and its associated cost vary depending on different deployment scenarios along with viability of the deployment scenarios, technical challenges, drivers of MPTCP adoption and benefits. According to \cite{Tselentis}, MPTCP is highly feasible in a scenario that delivers added incentives to the users or where the Multi-Homing Agent (see in Fig.~\ref{fig_NR_4}) prevails already for different reasons. From a technical point of view, MPTCP requires only a reasonably small modification to the TCP/IP stack at the end hosts (e.g., UE, MN/MeNB, SGW). There are implicitly two different costs. The first cost is involved with MPTCP software deployment (e.g., installing a particular extension or bundling with the operating system) and the second cost is involved with the additional wireless connectivity. A typical mobile user may not be interested in MPTCP if it entails extra costs. However, the MNOs can effort to deploy MPTCP to enjoy the MPTCP benefits. On the other hand, the UE and network equipment vendors/manufacturers can enjoy being a market differentiator by adding MPTCP features.

\section{Performance Analysis}
Robustness is a very critical area of concern in network design, especially in wireless access systems. In the event of an MN/SN link outage and/or a PDN connection failure/malfunctioning, using DC over MPTCP, the UE still has access to the network via the MN-UE/SN-UE link and/or the other remaining PDN connection as discussed in Sec.~\ref{MADC}. DC over MPTCP enables an exchange of data  between the MPTCP end-points in two  distinctive ways. The first one is dynamic distribution ( both alternating and simultaneous) of the traffic over the available paths, and the second one is a duplicate transmission of same data over the available paths. 

\begin{figure*}
  \centering
   \includegraphics[scale=.50]{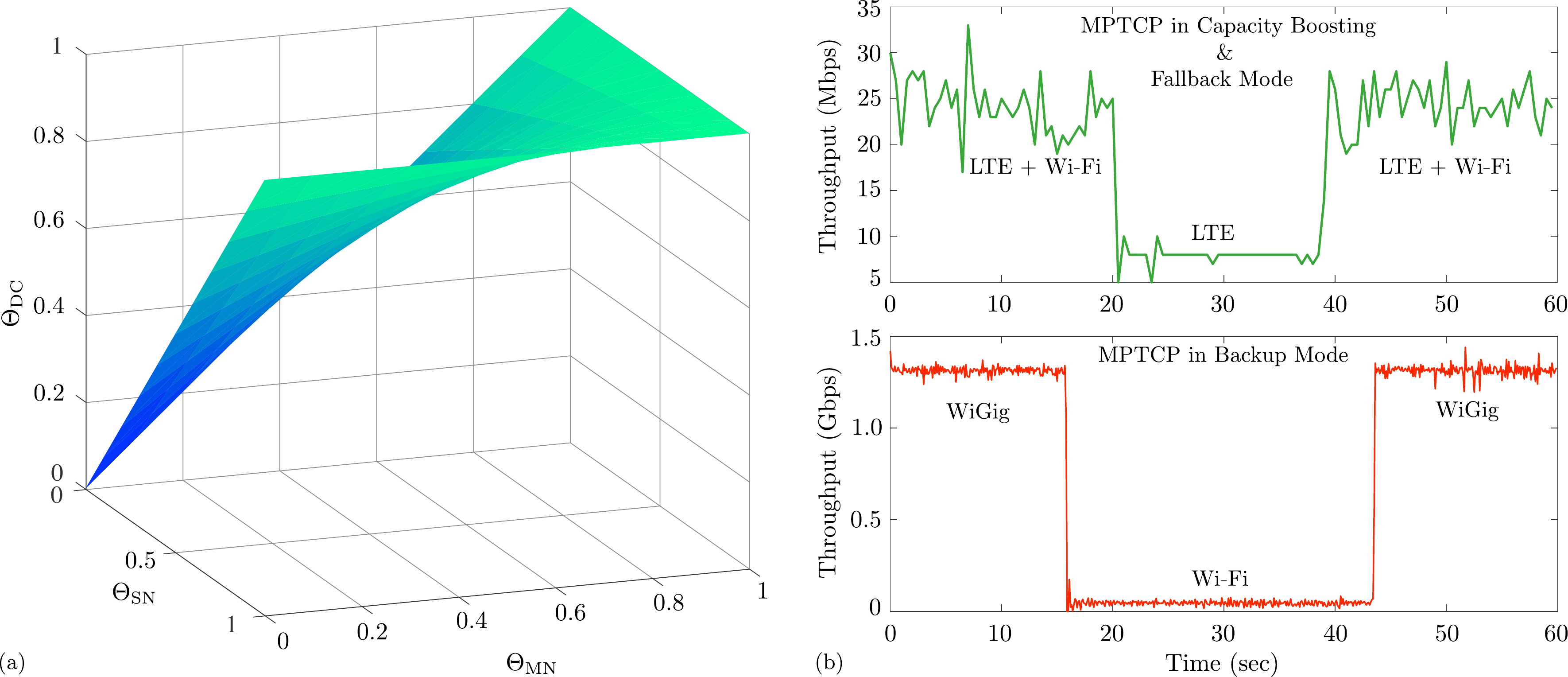}
   \caption{a). Availability of a communication channel when multiple paths are available. (b) up: MPTCP over LTE/Wi-Fi links in carrier aggregation mode. down: MPTCP in backup mode under non-3GPP access scenario with WiGig and Wi-Fi.  In both cases, we use the popular tool iperf3 for generating and collecting TCP and user datagram protocol traffic flows. The in-flight throughput value of iperf3 is tracked every 0.5s.}
   \label{fig_NR_5}
\end{figure*}

We evaluate the robustness achievable through DC over MPTCP considering the OpenFlow principle. The robustness of OpenFlow channel can be represented via the availability of multiple paths when the same traffic is simultaneously communicated over dual paths available for DC. Let $\Theta_{\rm MN}$ (between MPTCP instantiation point and MN) and $\Theta_{\rm SN}$ (between MPTCP instantiation point and SN) be the average availability of the paths (e.g., the fraction of time the interface works or establishes a successful communication link between the source and destination). To enhance the robustness, the data is simultaneously transmitted over both the available paths. Understandably, communication link fails only if both paths fail, which is represented by
\begin{equation}
\Theta_{\rm DC}=1-(1-\Theta_{\rm MN})(1- \Theta_{\rm SN}),\hspace{1mm}\Theta_{\alpha}=\frac{\psi_\alpha}{\psi_\alpha+\gamma_\alpha},
\end{equation}
where $\psi_{\alpha}$ and $\gamma_{\alpha}$ are the mean values of the uptime and downtime of the available path $\alpha\in\{{\rm MN,SN}\}$. 
It is obvious that the larger the value of availability the higher is the value of robustness, which is reflected in Fig.~\ref{fig_NR_5}a. Since the robustness of communication link is directly connected to the availability of the paths, MPTCP with DC can improve the legacy DC performance. As already mentioned earlier, bearer duplication over two different paths facilitates reliable communication through the DC feature. 

We, thereafter, investigate the efficacy of MPTCP in boosting throughput and in making the data transmission robust against the link failures by realizing the aggregation and fallback mechanism, respectively, between two paths in a scenario where the LTE link operates as the regular path. Fig.~\ref{fig_NR_5} (b): up clearly shows the ability of MPTCP in using the multiple available paths simultaneously. When the Wi-Fi service becomes available to the UE, the UE throughput is boosted. If the Wi-Fi link fails or the service becomes unavailable, MPTCP has the inherent capability to swiftly fall back to its default LTE connection, i.e., single-path TCP. Furthermore, MPTCP also supports the backup mode operation for reliable communication where there exists a standby link (e.g., Wi-Fi) as shown in Fig.~\ref{fig_NR_5} (b): down. The standby link becomes active only when the main link (e.g., WiGig) fails or becomes unavailable. We can observe that the TCP traffic is seamlessly and automatically switched from the WiGig to Wi-Fi when the WiGig link fails. More importantly, the WiGig link gets automatically reactivated when it becomes available again. Therefore, implementability of reliable communication employing MPTCP is confirmed.

\section{Conclusion}
This article gives an overview of the NR based DC feature as being standardized in 3GPP. We focus in particular on the standardization activities within the 3GPP related to small cell enhancement through DC and its variants in 5G NR. The architectural options, protocol stacks encompassing the migration from legacy LTE to full-fledged 5G and the deployment scenarios have been discussed in detail. We shed some light on integration of MPTCP with 3GPP DC and multi-access connectivity to bring in the advantages of MPTCP in terms of reliability and dynamic mapping between the traffic flows and the available paths.

Future Directions: Standardization organizations are actively putting enormous efforts in meeting the diversified requirements set by the current and future use-cases and applications. Integrating terrestrial (e.g., 5G NR) wireless with satellite systems for ubiquitous always-on broadband access everywhere, has been an area where a lot of efforts are being dedicated. The next-generation wireless networks, i.e., 5G, beyond 5G are evolving into very complex systems because of the very diversified service requirements, heterogeneity in applications, devices, and networks. The need to automate various functions of the networks has been one of the important requirements in order to reduce the operational expenses. More standardization efforts will be devoted to making the network self-organized and intelligent based on data analytics. The wireless community is opting for a highly energy efficient next-generation wireless, where the energy consumption should not be larger than that of today's networks, while still fulfilling 1000 times capacity gain. A lot of efforts will be put towards this direction.

\section*{Acknowledgement}
This research was conducted under a contract of R\&D for Expansion of Radio Wave Resources, organized by the Ministry of Internal Affairs and Communications, Japan.

\end{document}